\documentclass[submission,copyright,creativecommons]{eptcs}
\providecommand{\event}{WORDS 2011} \usepackage{breakurl}

\usepackage{amsfonts}
\usepackage{amsthm}
\usepackage{amsmath}

\def \ind{_{n \in {\mbox{\rm {\scriptsize I$\!$N}}}}}

\newcommand{\GN}{\mbox{\rm I$\!$N}}

\newcommand{\gN}{\mbox{\rm \scriptsize I$\!$N}}

\newcommand{\ab}{|}

\newtheorem{theorem}{Theorem}[section]
\newtheorem{lemma}[theorem]{Lemma}

\newtheorem{proposition}[theorem]{Proposition}

\newtheorem{definition}{Definition}[section]

\title{Dynamical generalizations of the Lagrange spectrum}

\author{S\'ebastien Ferenczi
\institute{Institut de Math\'ematiques de Luminy, CNRS - UMR 6206\\
Case 907, 163 av. de Luminy, F13288 Marseille Cedex 9 (France)}
\institute{F\'ed\'eration de Recherche des Unit\'es de Math\'ematiques de Marseille, CNRS - FR 2291}
\email{ferenczi@iml.univ-mrs.fr}}

\def\titlerunning{Generalizations of Lagrange spectrum}
\def\authorrunning{S. Ferenczi}

\begin{document}

\maketitle

\begin{abstract} 
We compute two invariants of topological conjugacy, the upper and lower limits of the inverse of Boshernitzan's $ne_n$, 
where $e_n$ is the smallest measure of a cylinder of length $n$, for three families of symbolic systems, the natural codings 
of rotations and three-interval exchanges and the Arnoux-Rauzy systems. The sets of values of these invariants for a given family of 
systems generalize the Lagrange spectrum, which is what we get for the family of rotations with the upper limit of $\frac{1}{ne_n}$.
\end{abstract}

The {\em Lagrange spectrum} is the set of  finite values of $L(\alpha)$ for all irrational numbers $\alpha$, where $L(\alpha)$ is the largest constant $c$ such that $\ab \alpha -\frac{p}{q}\ab \leq \frac{1}{cq^2}$ for infinitely many integers $p$ and $q$. It was recently remarked that this arithmetic definition can be replaced by a definition involving symbolic dynamics through the Sturmian sequences, which arise from natural coding of the irrational rotations of angle $\alpha$ by the  partition $\{[0,1-\alpha[, [1-\alpha, 1[\}$. Namely, as we prove in Theorem \ref{pas} below, which was never written before, $L(\alpha)$ is also the upper limit of the inverse of the so-called {\em Boshernitzan's $ne_n$}, where $e_n$ is the smallest (Lebesgue) measure of the cylinders of length $n$. 

Thus, for any symbolic dynamical system, it is interesting to compute two new invariants of topological conjugacy, $\limsup_{n \to+\infty}\frac{1}{ne_n}$ and $\liminf_{n \to+\infty}\frac{1}{ne_n}$. Moreover, for a given family of systems, the set of all values of these invariants can be called the {\em upper}, resp. {\em lower BL} (for Boshernitzan and Lagrange) {\em spectrum}. In this paper, we compute these spectra for three families of systems:  the irrational  rotations (seen as two-interval exchanges), the three-interval exchanges, both coded by the natural partition of the interval generated by the discontinuities, and the Arnoux-Rauzy systems. In each of these cases, we use an induction (or renormalization) process, which is respectively a variant of the Euclid algoritm, the self-dual induction of \cite{fz1}, and the natural one defined in \cite{ar}. A multiplicative form of the process yields explicit formulas for our invariants, and these formulas are then exploited in each case by using the underlying algorithm of approximation of real numbers by rationals, which is respectively the classical continued fraction expansion, an extension of a semi-regular continued fraction expansion, and the algorithm which motivated the study of Arnoux-Rauzy systems.

What we get in the end is a first partial description of the five new sets we introduced beside the classical Lagrange spectrum. For rotations, the lower BL spectrum is a compact set starting with $1$ and  an interval  (at least) as far as $1,03...$, ending at $1,38...$, with gaps,  above an accumulation point at $1,23...$. For three-interval exchanges, the upper BL spectrum looks, perhaps deceptively, like two times the Lagrange spectrum, starting at $2\sqrt{5}$ with gaps and an accumulation point at $6$, and ending with an interval  (at least) from $14,8...$ to infinity; the lower BL spectrum is fully determined and is none other than the interval $[2,+\infty]$. For Arnoux-Rauzy systems, we deal with cubic numbers and our knowledge is only embryonic: the upper BL spectrum starts at $8,44...$, with gaps, and ends at infinity, the lower BL spectrum starts at $2$ and  ends at infinity.

As a consequence, we get new uniquely ergodic systems for which $ne_n$ does not tend to zero when $n$ tends to infinity, showing that Boshernitzan's criterion is not a necessary condition; their existence in the family of three-interval exchanges was known but a proof was never written, while the examples in the family of Arnoux-Rauzy systems are new, and  surprising as these systems are often thought to behave like rotations, see the discussion at the end of Section 4.

\section{Preliminaries}
For a full study of the Lagrange spectrum, we refer the reader to the monograph \cite{cufl}; the following definition is equivalent to the one given in the introduction of the present paper.

\begin{definition} The {\em Lagrange spectrum} is  the set of all finite values of $$\limsup_{k\to +\infty}\frac{1}{q_k\left\ab q_k\alpha-p_k\right\ab},$$ for $\alpha$ irrational, the $\frac{p_n}{q_n}$ being the convergents of $\alpha$ for the Euclid algorithm.\end{definition} 

Let us just recall that the Lagrange spectrum is a closed set, its lowest elements are $\sqrt{5}$, then $2\sqrt{2}$, and discrete values up to a first accumulation point at $3$; above $3$, its structure is more complicated and not yet fully known, but it contains every real number above a value (which is known to be optimal) near $4,52...$ 

Note that to get the {\em Markov spectrum}, we replace the upper limit by a supremum in the above definition; the Markov spectrum will not be used in the present paper.

\begin{definition} The {\em symbolic dynamical system} associated to a language $L$ is the one-sided shift $$S(x_0x_1x_2...)=x_1x_2...$$ on the subset $X_L$ of ${\mathcal A}^{\gN}$ made with the infinite sequences such that for every $r<s$, $x_r...x_s$ is in $L$.

\noindent For a word $w=w_1...w_r$ in $L$, the {\em cylinder} $[w]$ is the set $\{x\in X_L; x_0=w_1, ..., x_{r-1}=w_r\}$. 

\noindent $(X_L,S)$ is {\em minimal} if $L$ is uniformly recurrent.

\noindent $(X_L,S)$ is {\em uniquely ergodic} if there is one  $S$-invariant probability measure  $\mu$; then  the {\em frequency} of the word $w$ is the measure $\mu[w]$. \end{definition}

\begin{definition}\label{sy} For a transformation $T$ defined on a set $X$, partitioned into $X_1$, ... $X_r$,  and a point $x$ in $X$, its {\em trajectory} is the infinite
sequence
$(x_{n})\ind$ defined by $x_{n}=i$ if $T^nx$ falls into
$X_i$, $1\leq i\leq r$. 

\noindent The language $L(T)$ is the set of all finite factors of its trajectories.

\noindent The {\em coding} of $(X,T)$ by the partition  $\{X_1,...X_r\}$ is the symbolic dynamical system $(X_{L(T)},S)$. \end{definition}

In \cite{bos} M. Boshernitzan introduced the following quantity:

\begin{definition} Let $(X_L,S)$ be a minimal symbolic system. If $\mu$ is an $S$-invariant probability measure, for each natural integer $n$, we denote by $e_n(\mu)$ the smallest positive frequency of the words of length $n$ of $L$. If $\mu$ is the only invariant probability measure, $e_n(\mu)$ is simply denoted by $e_n$.\end{definition}

After partial results in \cite{bos} and \cite{v0}, it was proved in \cite{bos2} that whenever, for some invariant probability measure $\mu$,  $ne_n(\mu)$ does not tend to $0$ when $n$ tends to $+\infty$, then the system $(X_L,S)$ is uniquely ergodic. This sufficient condition for unique ergodicity has been known since \cite{v0} as {\em Boshernitzan's citerion}.\\

In the present paper, all systems considered are uniquely ergodic, and we consider the quantity $ne_n$ for its own sake. Thus we define

\begin{definition} $${\mathcal B}=\limsup_{n\to+\infty}\frac{1}{ne_n}, \quad {\mathcal B}'=\liminf_{n\to +\infty}\frac{1}{ne_n}.$$ \end{definition}

\begin{proposition} ${\mathcal B}$ and ${\mathcal B}'$ are invariants of topological conjugacy among uniquely ergodic symbolic dynamical systems. \end{proposition}

A first crude estimate can be given using the complexity function, \begin{lemma} \label{com} ${\mathcal B} \geq \limsup_{n\to+\infty}\frac{p_L(n)}{n}$, ${\mathcal B}'\geq \liminf_{n\to+\infty}\frac{p_L(n)}{n}$.\end{lemma}

This is enough  to show that Boshernitzan's criterion is not a necessary condition: there are uniquely ergodic symbolic systems of exponential complexity \cite{gri}, and thus with $ne_n\to 0$, see also the discussion at the end of Section 4. But of course the above lemma implies that the study of these invariants is  interesting only for systems of linear complexity, for which the question of necessity can be asked again.\\

In view of Theorem \ref{lag} below, we are led to define the following sets:

 \begin{definition} For a family of uniquely ergodic symbolic dynamical systems $(X_a,S)$, $a\in {\mathcal F}$, the {\em upper BL spectrum} is the set of all values of ${\mathcal B}$ taken by the systems in this family, and the {\em lower BL spectrum} is the set of all values of ${\mathcal B}'$ taken by the systems in this family. \end{definition}

\section{Rotations and the dynamical definition of the Lagrange spectrum}
Surely there is nothing new to find about irrational rotations? The computation of ${\mathcal B}$ in
Thorem \ref{pas} below, and the subsequent Theorem \ref{lag}, which was the main motivation for the present paper,  were known to P.  Hubert and T. Monteil (private communications), but never written to our knowledge. The quantity $\frac{1}{\mathcal B'}$ was indeed computed in \cite{che} (see also \cite{bcf}) as, for irrational rotations, it is equal to another invariant of topological conjugacy, the {\em covering number by intervals} \cite{che}, which involves covering the space by Rokhlin towers; the spectrum of its possible values is the object of a question in \cite{che} and in \cite{cas}, to which Theorem \ref{nat} below gives a first (to our knowledge), though belated and partial, answer. \\

Let $\alpha<\frac{1}{2}$ be an irrational number; the rotations with $\alpha>\frac{1}{2}$ are treated in a similar way and all the results in this section from Theorem \ref{pas} onwards remain valid; the rotation of angle $\alpha$, is also the two-interval exchange defined by
$$ Tx = \begin{cases}
    x + \alpha & \text{if } x \in X_1=[0, 1-\alpha[\\
    x - 1 + \alpha & \text{if } x \in X_2=[1- \alpha, 1[ .
  \end{cases}$$

With this definition, a rotation admits a natural coding, by the partition of $X=[0,1[$ into   $X_1$ and $X_2$. Then  $L(T)$ has complexity $n+1$ and the trajectories are called {\it Sturmian sequences}. Irrational rotations are minimal and uniquely ergodic.\\

To get Theorem \ref{pas} below, we rely on a computation of both frequencies and lengths of factors of Sturmian sequences, which was done in \cite{ber}, but which we provide again by using a different version of the classic Euclid algorithm, making the computations quicker and ready to be generalized.

\begin{theorem}\label{pas} For a rotation of irrational angle $\alpha=[0,b_1,..,.b_n,...]$, if we define
$v_k=[0,b_k,b_{k-1},... b_1]$ and $t_k=[0,b_{k+1},b_{k+2},...]$ then 
$${\mathcal B}=\limsup_{k\to+\infty}\left(\frac{1}{
v_k}+t_k \right)=\limsup_{k\to+\infty}\left(b_k+v_{k-1}+t_k \right), \quad {\mathcal B}'=\liminf_{k\to+\infty}\left(1+t_k
v_k\right).$$
\end{theorem}

\begin{theorem}\label{lag}The upper BL spectrum of the family of rotations is the union of the Lagrange spectrum and $+\infty.$
\end{theorem}

 As for the lower LB spectrum, it seems to have never been studied to our knowledge, and its study looks to be of the same level of difficulty as for the Lagrange spectrum. We give now some of the first results about it. Note that ${\mathcal B}'$ is not the lower limit of $\frac{1}{q_k\left\ab q_k\alpha-p_k\right\ab}$ and thus is not directly linked to the quality of the approximation of $\alpha$ by rationals.

\begin{theorem}\label{nat} The lower BL spectrum of the family of rotations has $1$ as its smallest element, with ${\mathcal B}'=1$ if and only if the angle has unbounded partial quotients. It is a closed set.\\
Its two largest elements are $\frac{5-\sqrt{5}}{2}=1,38196...$ and $3-\sqrt{3}=1,26794...$, and there is no other element above $\frac{5}{4}$. \\
It contains an accumulation point equal to $\sqrt{5}-1=1,2360...$\\
It contains the interval $[1,1+\frac{4}{83+18\sqrt{2}}=1,03688...]$.
\end{theorem}

The third highest number in this spectrum is $\frac{16-4\sqrt{6}}{5}=1,2404...$, as can be seen with longer computations; the point $\sqrt{5}-1$ is the highest accumulation point, but to prove it requires a machinery similar to the one used to prove Theorem 5 in Chapter 1 of \cite{cufl}.

\section{Three-interval exchanges}
\subsection{The transformations}
\begin{definition}
  Given two numbers
$0 < \alpha$, $0 < \beta$ with $\alpha +\beta<1$, we define a
{\em three-interval exchange} on $X=[0, 1[$ by
$$Tx = \begin{cases}
    x + 1 - \alpha & \text{if } x \in X_1=[0, \alpha[\\
    x + 1 - 2 \alpha - \beta & \text{if } x \in X_2=[\alpha, \alpha + \beta[\\
    x - \alpha - \beta & \text{if } x \in X_3=[\alpha + \beta, 1[ .
  \end{cases}$$
 \end{definition}

Throughout this section, we ask that $\alpha$ and $\beta$ satisfy the {\it  i.d.o.c condition}
of Keane, which means in that case that they do not satisfy any rational relation
of the forms $p\alpha+q\beta=p-q$, $p\alpha+q\beta=p-q+1$, or
$p\alpha+q\beta=p-q-1$, for $p$ and $q$ integers. 

The points $\alpha$ and $\alpha+\beta$ are the
discontinuities of $T$, while $\beta_1=1-\alpha-\beta$ and
$\beta_2=1-\alpha$ are the discontinuities of $T^{-1}$. The
i.d.o.c. condition ensures that the negative orbits of the discontinuities of $T$
are infinite and have an empty intersection (it is its original
definition; see \cite{fhz2} for the equivalence with the one stated
here).\\

A three-interval exchange admits a natural coding, by the partition of $X$ into   $X_1,X_2,X_3$. Under the i.d.o.c. condition, $(X,T)$ is minimal and uniquely ergodic and $L(T)$ has complexity $2n+1$.\\

Throughout this section, we add the conditions $0 < \alpha <
\frac{1}{2}$, and $2 \alpha + \beta > 1$; they ensure that the induction process described below does not have an irregular behaviour in the early stages: as is shown in \cite{fz1}, their absence modifies only a finite number of stages, and all the results in this section from Theorem \ref{len} onwards remain valid without these extra conditions.\\

\begin{theorem}\label{len} The smallest element in the upper BL spectrum of  three-interval exchange transformations, and the only one below $\frac{12+29\sqrt{3}}{13}=4,786...$, is $2\sqrt{5}=4,47...$.\\
The spectrum is a closed set and contains an accumulation point equal to $6$. \end{theorem}

The values of ${\mathcal B}$ between $2\sqrt{5}$ and $6$ were found only by trial and error; the second value is very likely to be $4\sqrt{2}=5,65... $,. The third value we found is $\frac{2\sqrt{2600}}{17}=5,9988...$, and the fourth one is $\frac{4\sqrt{209306}}{305}=5,999996...$.
 
 Thus the first, second, third, fourth smallest element we found in the upper BL spectrum of three-interval exchange transformations is respectively twice the first, second, sixth and twelfth smallest element in the  Lagrange spectrum. Though of course we might have missed some values, it seems likely that {\em the upper BL spectrum of three-interval exchanges below $6$ is strictly included in twice the Lagrange spectrum below $3$}; thus we conjecture that {\em $6$ is the lowest accumulation point of our spectrum}.

\begin{theorem}The upper  BL spectrum of the family of three-interval exchanges contains the interval $[12+2\sqrt{2}=14,828..., +\infty].$\end{theorem}

\begin{theorem} The lower BL spectrum of the family of three-interval exchanges is the interval $[2, +\infty].$ \end{theorem}

Thus for some uniquely ergodic three-interval exchange transformations we have $ne_n\to 0$ when $n$ tends to infinity; this result, and its consequence that Boshernitzan's criterion is not a necessary condition in this family of systems, are stated without proof in \cite{v}. Note that the covering number by intervals (see the opening of Section 2 above) of a three-interval exchange is shown in \cite{bcf} to be the same as for the inducing rotation, and thus is not equal to $\frac{1}{\mathcal B'},$ in contrast with the case of rotations.

\section{Arnoux-Rauzy systems}
The {\em Arnoux-Rauzy systems}  are defined in \cite{ar} as the minimal  symbolic systems on the alphabet $\{1,2,3\}$ such that the complexity of the language is $2n+1$ for all $n$, and, for all $n$, there are one right special and one left special word. Then \cite{ar} proceeds to give a constructive (additive) algorithm to generate them with
three families of words, built with three rules denoted by $a$, $b$ and $c$; \cite{cfme} gives a multiplicative version of this construction, which we take here as a definition, valid up to permutations of $\{1,2,3\}$: the $k_n$ are the number of consecutive times a given rule is used, while the $n_i>1$ mark the times where three consecutive rules are all different, such as, up to permutations of $\{a,b,c\}$, rule $a$ used $k_{n_i-1}$ times, then rule $b$ used $k_{n_i}$ times, then rule $c$ used $k_{n_i+1}$ times. 

\begin{definition}\label{arm} Given two infinite  sequences of integers  $k_n\geq 1$, $n\geq 1$, and $n_1<n_2...<n_i<...$
the Arnoux-Rauzy system $(X_L,S)$ defined by them is the symbolic system associated to the language $L$ of all factors of $(H_n)\ind$, where the three words $H_n$, $G_n$, $J_n$ are built from $H_0=1$, $G_0=2$, $J_0=3$ by two families of rules:
\begin{itemize} \item if $n+1=n_i$ for some $i$,
 $H_{n+1}=G_nH_n^{k_{n+1}}$, $G_{n+1}=J_nH_n^{k_{n+1}}$, $J_{n+1}=H_n$; 
  \item otherwise, $H_{n+1}=G_nH_n^{k_{n+1}}$, $G_{n+1}=H_n$,
  $J_{n+1}=J_nH_n^{k_{n+1}}$.\end{itemize}\end{definition}

Every Arnoux-Rauzy system is minimal \cite{ar} and uniquely ergodic (by  \cite{bos3} because the complexity is $2n+1$). Though they are defined as symbolic systems, they have also geometric models, see \cite{a} \cite{abi} \cite{ar}\cite{rau}: every Arnoux-Rauzy system is a coding of a six-interval exchange on the circle, and some of them are codings of rotations of the $2$-torus.

\begin{proposition} The upper BL spectrum of the family of Arnoux-Rauzy systems contains $+\infty$, which is reached if and only if  the $k_n$, $n\in\GN$, or the  $n_{i+1}-n_i$, $i\geq 1$, are unbounded. Its smallest element, and the only one below $\frac{181}{21}=8,619...$,  is reached for the {\em Tribonacci system} where $n_i=i$ for all $i\geq 1$ and $k_n=1$ for all $n\geq 1$; for this system, if $y=1,8392...$ is the root bigger than $1$ of the polynomial $X^3-X^2-X-1$, then ${\mathcal B}=2y^2+\frac{4y}{y^2+1}=8,4445...$. \end{proposition}

\begin{theorem}\label{lar} The smallest element in the lower BL spectrum of the family of Arnoux-Rauzy systems is $2$, and the largest  is $+\infty$. Every integer  greater or equal to $2$ is in the lower BL spectrum, and is an accumulation point, as is $+\infty$.  \end{theorem}
  
Thus we have a new, and very simple,  family of counter-examples to the necessity of Boshernitzan's criterion. 

Of course, there are other values in the lower spectrum than those in Theorem \ref{lar}, and we conjecture that {\em the lower BL spectrum of the family of Arnoux-Rauzy systems is the interval $[2,+\infty]$.}\\

The Arnoux-Rauzy systems raise questions about rotations of the $2$-torus, and we may ask what could be the BL spectra for that family of systems, but the problem is that they do not admit any coding which may be called natural. If we code a rotation of the $2$-torus  with the Cartesian product of two partitions of the $1$-torus, then the complexity is quadratic and all ${\mathcal B}$ and ${\mathcal B}'$ are infinite by Lemma \ref{com}, which gives another trivial counter-example to the necessity of Boshernitzan's criterion, but one can object that it just means the coding is not appropriate. The Arnoux-Rauzy systems were devised to provide codings with linear complexity for rotations of the $2$-torus, but this was succesful only in a limited number of cases. Still, if we consider these cases, the tentative lower BL spectrum of the family of rotations of the $2$-torus seems to be quite different from the lower BL spectrum of rotations of the $1$-torus: if we take an Arnoux-Rauzy system with $n_i=i$ and constant $k_n=k$, it is a coding of a rotation of the $2$-torus by \cite{abi}, and these give arbitrarily high values for ${\mathcal B}'$; if $n_i=i$ and $k_n$ grows slowly (for example $k_n\leq \frac{1}{15}n$), we get an infinite ${\mathcal B}'$ while the Arnoux-Rauzy system is shown in \cite{cfme} to have two continuous eigenfunctions, and is still conjectured to be a coding of a rotation of the $2$-torus.

\bibliographystyle{eptcs}
\bibliography{lagwords1}

\end{document}